\documentclass[5p,twocolumn]{elsarticle}
\journal{Nuclear Instruments and Methods in Physics Research Section A}
\usepackage[USenglish]{babel}
\usepackage[T1]{fontenc}
\usepackage[utf8]{inputenc}
\usepackage{amssymb} 
\usepackage{amsmath}
\usepackage{amsthm}
\usepackage{amsfonts}
\usepackage{siunitx}
\usepackage{booktabs}

\begin{document}

%%%%%%%%%%%%%%%%%%%%%%%%%%%%%Frontmatter%%%%%%%%%%%%%%%%%%%%%%%%%%%%%%%%%

\title{Simulation of deflecting structures for dielectric laser driven accelerators}

\author[desy,uhh]{W.~Kuropka}
\ead{willi.kuropka@desy.de}

\author[desy]{R.~Aßmann}
\author[desy]{U.~Dorda}
\author[desy,uhh]{F.~Mayet}

\address[desy]{Deutsches Elektronen-Synchrotron DESY, Notkestraße 85, 22607 Hamburg, Germany}
\address[uhh]{Universität Hamburg, Mittelweg 177, 20148 Hamburg, Germany}

\begin{abstract}
In laser illuminated dielectric accelerators (DLA) high acceleration gradients can be achieved, due to high damage thresholds of the materials at optical frequencies. This is a necessity in developing more compact particle accelerator technologies. The Accelerator on a CHip International Program funded by the Gordon and Betty Moore Foundation is researching such devices. Means to manipulate the beam, i.e. focusing and deflection, are needed for the proper operation of such devices. These means should rely on the same technologies for manufacturing and powering like the accelerating structures. In this study different concepts for dielectric laser driven deflecting structures are investigated via particle-in-cell (PIC) simulations and compared afterwards. The comparison is conducted with respect to the suitability for beam manipulation. Another interesting application will be investigated as a diagnostic device for ultra short electron bunches from conventional accelerators functioning like a radio frequency transverse deflecting cavity (TDS).
\end{abstract}


\maketitle

%%%%%%%%%%%%%%%%%%%%%%%%%%%%%Mainmatter%%%%%%%%%%%%%%%%%%%%%%%%%%%%%%%%%

%%%%%%%%%%%%%%%%%%%%%%%%%%%%%SECTION%%%%%%%%%%%%%%%%%%%%%%%%%%%%%%%%%

\section{Introduction}

The Accelerator on a CHip International Program (ACHIP) is a research project funded by the Gordon and Betty Moore Foundation. It aims at the construction of a compact fully laser driven electron accelerator for radiation generation and atto-second science. Several universities in Europe and the USA and the national laboratories PSI, DESY and SLAC are involved. \cite{ACHIP}

This research field gains more attention in recent years as the search for compact particle acceleration technologies continues. Fully dielectric laser driven acceleration structures are foreseen to sustain high acceleration gradients in the \SI{}{\giga\volt\per\m} regime. This is mainly due to the high laser damage thresholds of dielectrics at optical frequencies. The advancement in micro- and nano-fabrication from the semi conductor industry can be leveraged in the manufacturing of these structures limiting production costs. Also the ongoing development in laser technologies is promising.\cite{England2014}

Structures for focusing and deflection of the beam are necessary for the eventual implementation of such an accelerator. Deflecting structures also can be useful diagnostics for short electron bunches. In this paper three candidates for the deflection of particles are investigated via PIC simulation.

%%%%%%%%%%%%%%%%%%%%%%%%%%%%%SECTION%%%%%%%%%%%%%%%%%%%%%%%%%%%%%%%%%

\section{Grating based dielectric laser acceleration}
\begin{figure}[!htb]
   \centering
   \includegraphics*[width=85mm]{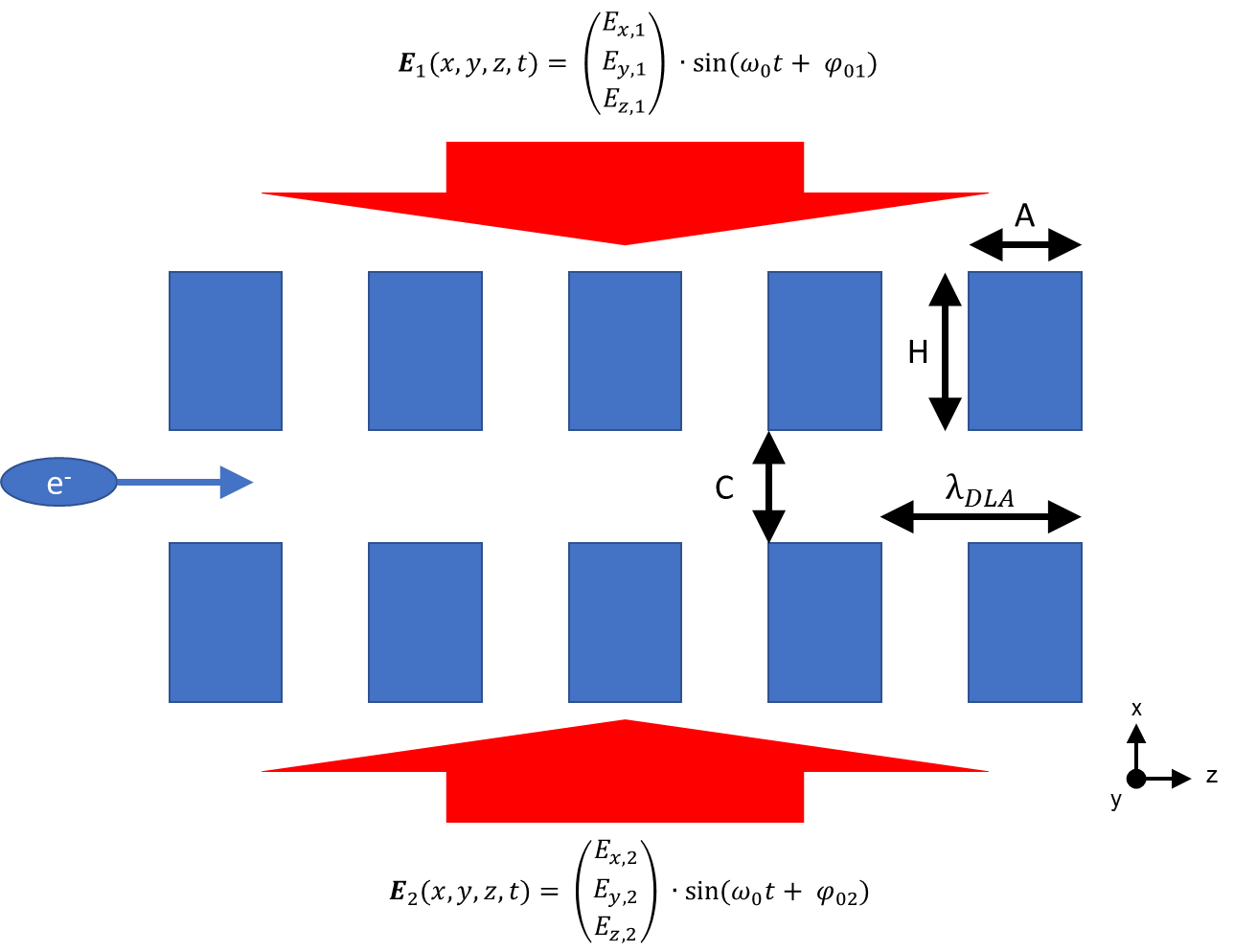}
   \caption{Schematic of the double illuminated pillar grating type DLA with parameters $\lambda_{DLA} = \SI{1.994}{\micro\m}$, $A = 0.5 * \lambda_{0}$, $C = 0.37 * \lambda_{0}$, $H = 0.87 * \lambda_{0}$ for the laser wavelength from \ref{tab:laserParams} with fused silica as dielectric}
   \label{fig:DLA_schematic}
\end{figure}

The investigated structure consists of two gratings with a gap. They are illuminated form both sides with laser beams. The laser beams are in phase and have the same amplitude. Electric fields of both beams are polarized in the direction the electron beam travels along the gap. Figure \ref{fig:DLA_schematic} is illustrating the structure. 

The periodic diffraction fields in the gap along the z-axis can be described via spatial harmonics. If the grating period matches the incoming laser wavelength, the first spatial harmonic has a speed of light phase velocity. The acceleration in this harmonic is transversely uniform depending on the gap size. The phase velocity of this first spatial harmonic can be reduced by shortening the grating period with respect to the laser wavelength making it possible to also accelerate non-relativistic particles. Alternatively higher order modes can be chosen to be synchronous with the particle according to this synchronicity condition:

\begin{equation}
	\lambda_{DLA} = n \beta \lambda_{laser}
	\label{eq:synchCond}
\end{equation}

Where n is the order of the spatial harmonic, $\beta$ is the fraction of the vacuum speed of light of the particle, $\lambda_{laser}$ is the central wavelength of the incoming laser field and $\lambda_{DLA}$ gives the period of the grating structure.

%%%%%%%%%%%%%%%%%%%%%%%%%%%%%SECTION%%%%%%%%%%%%%%%%%%%%%%%%%%%%%%%%%

\section{Grating based deflecting structures}

The presented structures are grating based due to the necessary compatibility with the grating type accelerator design. For acceleration the electric field of the incoming laser is linearly polarized in longitudinal direction of the electron beam.

In the first investigated scheme (A) the polarization is rotated $90^\circ$ into the y-direction transverse to the e-beam. The deflection now should occur in the “free” direction not limited by the grating aperture .

For the second scheme (B) the phase of the two incoming beams is set off by $180^\circ$ canceling out the accelerating fields and amplifying the transverse components. Here the deflection appears in the direction limited by the grating (See figure \ref{fig:DLA_schematic} and table \ref{tab:Polarization}).

\begin{figure}[!htb]
   \centering
   \includegraphics*[width=78mm]{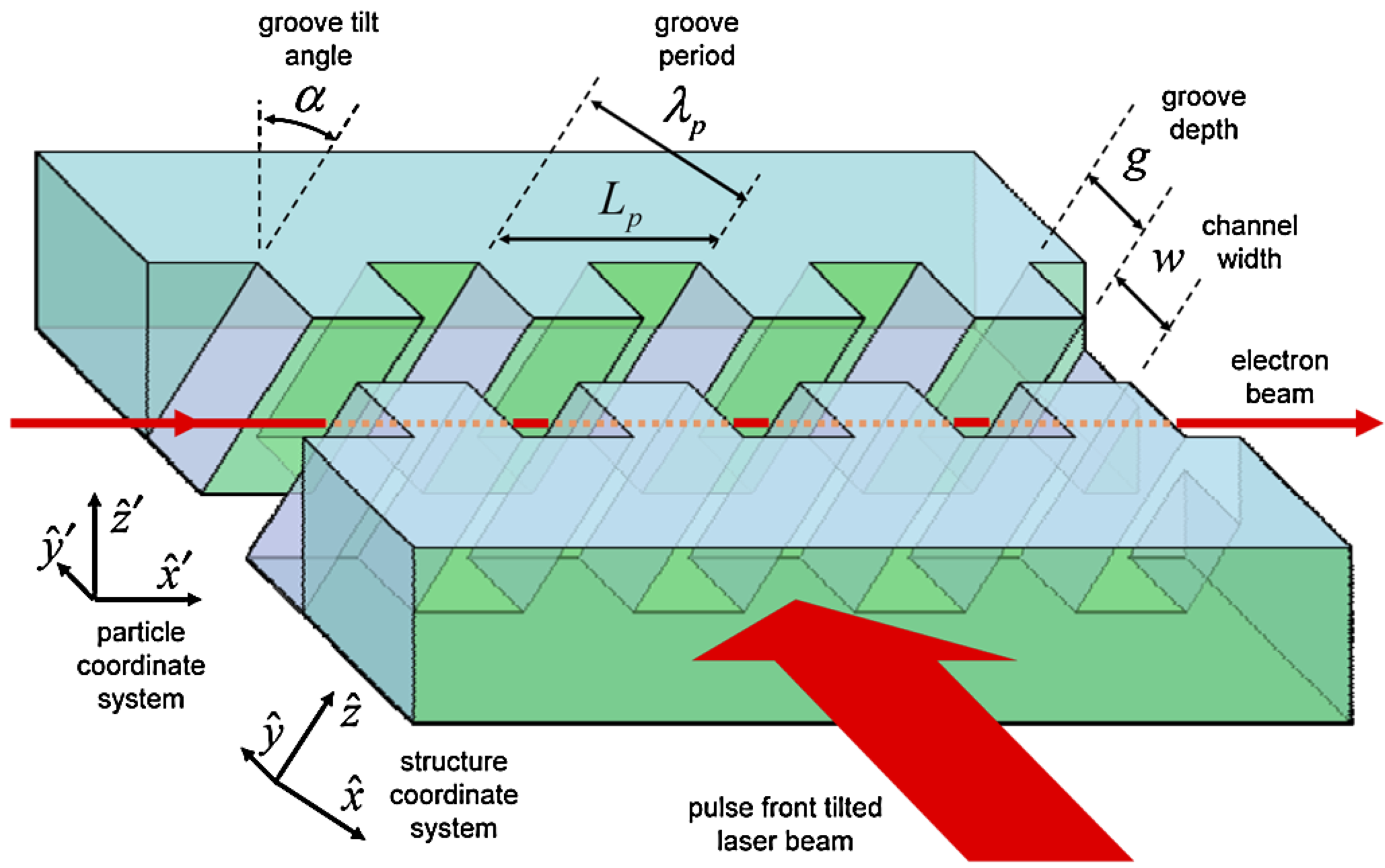}
   \caption{The whole structure is rotated so that the electron beam travels through the gap at an angle relative to the grating grooves. Illustration taken from \cite{Plettner} Scheme C.}
   \label{fig:DLA_TDS_C}
\end{figure}

The third scheme (C) is from \cite{Plettner} and a rotation of the whole acceleration structure so that the force has an additional strong transverse component as shown in figure \ref{fig:DLA_TDS_C}. The polarization of the electric field of the laser is still perpendicular to the direction of the grating groves. The other component is still in the longitudinal direction and will induce additional energy spread if used as streaking device.

\begin{table}[hbt]
   \centering
   \caption{Laser Parameters for Acceleration and Schemes A and B}
	 \vspace{0.1cm}
   \begin{tabular}{llll}
      \toprule
      \textbf{E-Field} & \textbf{Accelerator} & \textbf{Scheme A} & \textbf{Scheme B} \\
      \midrule
      $E_{x,1}$ & 0 & 0 & 0\\
      $E_{y,1}$ & 0 & $E(x,y,z,t)$ & 0\\
      $E_{z,1}$ & $E(x,y,z,t)$ & 0 & $E(x,y,z,t)$\\
			$E_{x,2}$ & 0 & 0 & 0\\
			$E_{y,2}$ & 0 & $E(x,y,z,t)$ & 0\\
			$E_{z,2}$ & $E(x,y,z,t)$& 0 & $E(x,y,z,t)$ \\
			$\phi_{01} - \phi_{01}$ & 0 & 0 & $180^\circ$\\
     \bottomrule
   \end{tabular}
   \label{tab:Polarization}
\end{table}

%beforehand pictures
%\begin{figure}[!htb]
   %\centering
   %\includegraphics*[width=78mm]{Images/DLA_Acc.png}
   %\caption{Laser configuration for using the structure as an accelerator.}
   %\label{fig:DLA_Acc}
%\end{figure}
%
%\begin{figure}[!htb]
   %\centering
   %\includegraphics*[width=78mm]{Images/DLA_TDS_A.png}
   %\caption{The Polarization of the incoming laser beams is rotated transversely to the electron beam direction. Scheme A.}
   %\label{fig:DLA_TDS_A}
%\end{figure}
%
%\begin{figure}[!htb]
   %\centering
   %\includegraphics*[width=78mm]{Images/DLA_TDS_B.png}
   %\caption{The phase of the incoming laser beams is offset by $180^\circ$. Scheme B.}
   %\label{fig:DLA_TDS_B}
%\end{figure}

%%%%%%%%%%%%%%%%%%%%%%%%%%%%%SECTION%%%%%%%%%%%%%%%%%%%%%%%%%%%%%%%%%

\section{PIC simulation}

The CST PIC \cite{CST} simulations were set up with the parameters from table \ref{tab:laserParams} for one grating period with a point source test beam and excited by plane wave. The deflection is laser-to-electron phase dependent and the maximum is shown in table \ref{tab:MAXs}. The dimensions of the grating used in the simulation are shown in figure \ref{fig:DLA_schematic}.

Additionally the transverse kick from the acceleration structure is investigated as comparison. It is important to note that this kick is position dependent and particles at the center of the gap are not kicked. This is comparable to a focusing or defocusing of an electron beam. Here the offset for the test beam is \SI{150}{\nm} from the gap center.

\begin{table}[hbt]
   \centering
   \caption{Simulation Parameters}
	 \vspace{0.1cm}
   \begin{tabular}{ll}
      \toprule
      \textbf{Parameter} & \textbf{Value}\\
      \midrule
      Wavelength & \SI{2}{\micro\meter}\\
      Laser Amplitude & \SI{2}{\giga\volt\per\meter}\\
      Particle Energy & \SI{5}{\mega\eV}\\
      \bottomrule
   \end{tabular}
   \label{tab:laserParams}
\end{table}

\begin{table}[hbt]
   \centering
   \caption{Simulation results}
	 \vspace{0.1cm}
   \begin{tabular}{lllll}
      \toprule
      \textbf{Scheme} & \textbf{A} & \textbf{B} & \textbf{C} & \textbf{Accel.}\\
      \midrule
      Max. defl. & \SI{4}{\micro\radian} & \SI{20}{\micro\radian} & \SI{34}{\micro\radian} & \SI{0.5}{\micro\radian}\\
      Defl. plane & unlimited & limited & unlimited & focusing\\
      \bottomrule
   \end{tabular}
   \label{tab:MAXs}
\end{table}

%%%%%%%%%%%%%%%%%%%%%%%%%%%%%SECTION%%%%%%%%%%%%%%%%%%%%%%%%%%%%%%%%%
	
\section{Results}

\begin{figure}[!htb]
   \centering
   \includegraphics*[width=78mm]{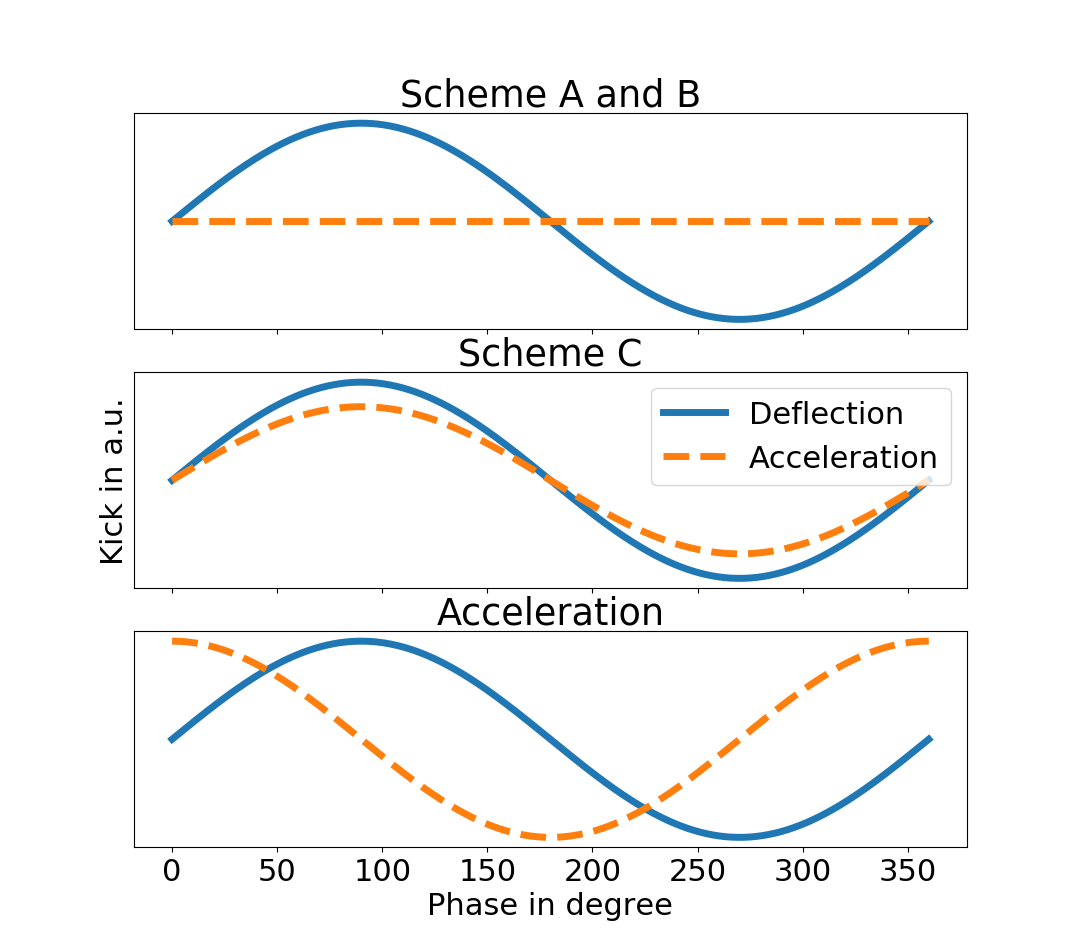}
   \caption{The top graph shows the dependecy of the transverse and longitudinal kicks for scheme A and B. There are no significant longitudinal kicks. The middle graph shows the relation for scheme B where the significant longitudinal an transverse kicks are in phase. The third graph shows the relation for the grating used as an accelerator.}
   \label{fig:DeflOverPhase}
\end{figure}

The results show that the schemes B and C may be feasible candidates for deflection or streaking devices. In scheme B the beam size is limited by the aperture of the structure which limits the resolution in a streaking application. For beam manipulation the scheme may still be feasible. Scheme C has the strongest deflection forces but inherently has longitudinal forces in the sam order of magnitude, too, which would add to the energy spread of the streaked particles. Both schemes have transverse forces an order of magnitude stronger than the focusing/defocusing forces of the acceleration structure.
Scheme A has only limited deflection forces. The electric field and the effect of the magnetic field are canceling each other out. In figure \ref{fig:DeflOverPhase} the relations for the different schemes are shown.

%%%%%%%%%%%%%%%%%%%%%%%%%%%%%SECTION%%%%%%%%%%%%%%%%%%%%%%%%%%%%%%%%%

\section{Outlook and Conclusion}

Due to the dependence of the deflecting force on the laser-to-electron phase and the short laser wavelength in use further investigation is necessary to study the beam dynamics in such a system. Results on that topic can be found in \cite{FrankEAAC2}. For use as a streaking device the maximum bunch length is limited to one quarter of the driving wavelength. With the equation from \cite{roehrs} for the longitudinal TDS resolution for electrons at optimal phase advance to the detector screen

\begin{equation}
\sigma_{long} = \frac{\sqrt{\epsilon_y} c^2 \lvert p \rvert}{\sqrt{\beta_y} 2 \pi f_0 e V_0}
\label{eq:TDSresolution}
\end{equation}

one can see that a large beta function at the TDS ($\beta_y$) is beneficial for the longitudinal resolution. The limited beta function due to the smaller apertures of DLA can be compensated by the high frequency ($f_0$) of the incoming fields. The high intensities achievable lead to a high streaking voltage $V_0$. The remaining quantities are the transverse emittance $\epsilon_y$, the elementary charge $e$ and the particle momentum $p$. Schemes A and C streak in the direction of the DLA which is not limited by the gap size. Here a bigger beta function might be achievable. Within this limitations a longitudinal resolution in the attosecond regime would be theoretically possible with these grating type structures.

%%%%%%%%%%%%%%%%%%%%%%%%%%%%%SECTION%%%%%%%%%%%%%%%%%%%%%%%%%%%%%%%%%
	
\section{Acknowledgments}
	
This research was conducted under the Accelerator on a CHip International Program (ACHIP) funded by the Gordon and Betty Moore Foundation via grant GBMF4744.
	
%%%%%%%%%%%%%%%%%%%%%%%%BIB%%%%%%%%%%%%%%%%%%%%%%%%%%%%%%%%%%%%%%%%%%%%
\newpage
\section*{References}

\bibliographystyle{unsrtnat}
\bibliography{\jobname}
%\printbibliography

\end{document}